Live Captions in Virtual Reality (VR)

Pranav Pidathala, Dawson Franz, Dr. James Waller, Dr. Raja Kushalnagar, Dr. Christian Vogler

Corresponding author email: james.waller@gallaudet.edu

Gallaudet University

**Introduction**

Virtual Reality (VR) as a mainstream industry is an extremely recent development, with today's technology only becoming widely available around six years ago with the public release of VR headsets like the Oculus Rift, HTC Vive, and Playstation VR. Typically, in virtual reality, users have a 360º view of their surroundings: they can freely view everything in the virtual world by moving either their head or a controller. However, with novel technology comes novel accessibility challenges.

VR technology that uses spoken audio must provide that information in an alternative format to be accessible to Deaf and Hard of hearing (DHH) users – such as captions. Typically these appear at the bottom of a screen during the duration of the spoken audio, with standard guidelines such as a black background box, sans-serif white font, and a maximum of 3 lines with 40 characters per line.

However, virtual reality adds several new dimensions to consider. For example, when the person moves, should the captions move with them? Or do they stay fixed in place, anchored to some landmark in the virtual environment. The question of how best to design captions carries additional importance since a key advantage of virtual reality applications are their immersive-ness and poorly integrated elements could break this immersion. Previous work has identified several different approaches to caption positioning in VR, which can be divided into two approaches – 'headlocked' and 'dynamic' or 'world-locked (BBC Research Group, 2017). In the first approach, characterized as 'headlocked' captions, captions are consistently locked to a fixed position in the user's view. If the user shifts their head, the captions will follow, such that they stay in the same position in the user's view. In this way the captions are always visible to the user. One version of headlocked captions have a lag: the captions follow the head movements with a slight delay (allowing for smoother movement).

In contrast, in the 'world-locked' approach, the captions are not locked to a fixed position in the person's *view*, but rather are 'locked to the environment'. Just like other virtual objects, dynamic captions stay fixed in the environment as the person moves around. If the user turns away from the captions, they disappear from view; they do not follow the user's head movements. Some are 'speaker/source-locked' – meaning they appear near the source of the audio (usually below the speaker), or near an object associated with the caption content. Others are 'appear-locked' – locked to the user's initial head position – so when the captions appear, they appear where the user is looking, but unlike head-locked captions they stay in place if the user looks away.

A couple of previous approaches have compared different captioning styles with various users. Rothe et al. (2018) looks at hearing people's subtitle/caption preferences in a VR setting in which the audio is in a foreign language, comparing speaker-locked captions to head-locked captions (without lag). They found no significant preference between the two styles. Another study, Agulló (2019) tested two caption behaviors with 27 hearing and 13 DHH participants: headlocked captions without lag and 'fixed' (world-locked) captions. In the world-locked caption condition there were three different caption set ups spaced evenly around the user 120 degrees apart, such that wherever the user was looking there was likely at least one set of captions visible. They found that 87.5% of participants strongly preferred the headlocked captions over the multiple world-locked captions.

Our study tests three different VR caption types with DHH users: 1) headlocked captions without lag, 2) headlocked captions with lag, and 3) 'appear-locked captions' in which the captions always appear the user's visual field but stay locked to the world thereafter. Our study exclusively recruits from DHH users, who may have different preferences for VR captions – many DHH user are frequent caption users and unlike hearing users, they cannot necessarily rely on audio information to know if someone is speaking at that moment, and thus may use captions differently.

**Implementing the Virtual Reality environment and captions**

*Participants*

We recruited DHH participants from a university for deaf and hard of hearing students located in Washington, D.C. who used ASL and written English to communicate (and for some, spoken English as well).

*Materials*

We coded a testable prototype of three different captioning behaviors in Mozilla Hubs, a meeting platform for VR. We implement three caption conditions, as follows:

'Headlocked captions' (no lag)
- Always visible
- Movement on the x-axis, however it was locked in position on the y-axis: it cannot move up or down.

'Lag captions'
- Usually always visible, except when moving very fast.

- Do not move up to down. When moving on the x-axis, the captions usually take approximately one second to 'catch up' to the user's movement.

'Appear-locked captions'

- The caption box would be shown when the user is speaking. Once the speaker starts a new sentence, the captions would go to where the user is looking.
- If the user is looking in the same sport continuously, the captions would go under each other.

For consistency, all aspects of the captioning types other than their movement behavior are kept the same and use the same textbox object. Following Evans (), we emulate existing TV closed captioning guidelines for our caption textbox: white sans-serif text on a black box with up to three lines of text.

We used Mozilla Hubs as our meeting platform, and Javascript as our primary coding language. The website was hosted on AWS. We add onto Hubs' existing codebase using A-frame framework for 3D Objects with the React library. We populated the scene with various background elements, such as stairs and elevated platforms, to give the user an environment to visually inspect and explore as they travelled through.

Rather than precorded captions, we used Microsoft Azure's speech-to-text for the live voice transcription, with a W3C Speech Recognition Ponyfill so it works on all browsers. The goal is to mimic the using of VR for a live meeting or virtual tour.

*Procedure*

After the experimenter explains the instructions and controls, the participant puts on the VR headset, adjusts and calibrates the equipment, and familiarizes themselves with a practice environment.

During experiment, the participant experiences a 'live virtual tour'. The experimenter gives a scripted presentation about different products displayed in the VR environment, such that the participant could wander around the displays as the experimenter discussed each in turn. Automatically generated captions would appear in the participant's VR environment.

We did this three times, once for each caption condition. The participants also had the opportunity to provide open-ended feedback, and to pick their preferred caption behavior.

**Results**

8 DHH participants participated in our study. All participants answered that they use captions regularly. 6 participants use captions as their "primary method to understand information" and 2 use it as their secondary method. All participants had little to no previous experience with VR.

*User Preferences and Ratings*

Participants were split in their caption preferences: three preferred headlocked captions, two preferred lag captions, and two preferred the appear captions. In their comments, several participants said they liked that the headlocked and lag captions were consistently in the same place for reading the captions (both locked to head position) – though for others the captions were too low or too close, which was a source of frustration. Conversely those who liked appear captions liked that they moved relative to the head and did not stay in one position. Both participants who preferred the appear caption type use hearing devices and used captions as their secondary method of understanding information.

Overall, any issues that arose with the captions did not seem to turn participants off from the technology as a whole – post-study all but one participant reporting feeling extremely (4) or moderately (3) comfortable in VR, and several commented on enjoying the experience and feeling immersed.

**Discussion**

Users are not uniform in their preferences. For example, people who rely on captions as a secondary source of information may like to be able to 'look away' from captions, but others who rely primarily on captions appear to prefer headlocked captions that are always visible. Compared to TV closed captions, VR captions may require additional customization options and/or should exist as an interactable object, since the expanded view of VR may make a one-size-fit-all approach

Nonetheless, captioning work is promising, and shows that captions do not necessarily detract from the VR experience – many DHH participants are excited to participate in VR technology. So, it is important to keep these technologies accessible to all.